\documentclass[aps,prb,10pt,superscriptaddress,showpacs,amssymb,amsmath,twocolumn]{revtex4-1}
\usepackage{graphicx}
\usepackage{units}
\usepackage{color}

\newcommand{\buenosaires}{Departamento de F\'{\i}sica and IFIBA, 
FCEyN, Universidad de Buenos Aires, 
Ciudad Universitaria, Pab.\ I, C1428EHA Buenos Aires, Argentina}

\begin{document}

\title{Spin-orbit effects in the hydrogenic impurity levels of wurtzite semiconductors}

\author{O.\ L.\ Hern\'andez Rosero, J.\ I.\ Melo, and P.\ I.\ Tamborenea}
\affiliation{\buenosaires}


\begin{abstract}

The corrections to the $E_2^*$ energy level of hydrogenic impurities in semiconductors 
with wurtzite crystal structure are calculated using first-order perturbation theory in 
the envelope-function approximation.
We consider the intrinsic (Dresselhaus) spin-orbit effective Hamiltonian in the conduction 
band and compare its effects to the renormalized extrinsic (Rashba) spin-orbit interaction 
which is analogous to the spin-orbit interaction in the bare hydrogen atom.
In order to evaluate the extrinsic spin-orbit interaction we obtain the renormalized coupling
constant $\lambda^*$ for wurtzite semiconductors from 8-band Kane theory.
We apply our theory to four representative binary semiconductors with wurtzite crystal
structure, namely, GaN, ZnO, InN and AlN, and discuss the relative strength of the effects 
of the intrinsic and extrinsic spin-orbit contributions.

\end{abstract}

\maketitle


\section{Introduction}
\label{sec:introduction}

Hydrogenic impurities are a central aspect of semiconductor physics and technology.\cite{pan,yu-car}
In recent years, impurity states have been proposed as potential qubits in quantum information 
devices.\cite{kan,koi-hu-das}
In binary III-V and II-VI semiconductors, the spin-orbit interaction can play an important role
in the electronic structure of confined electronic states.\cite{win,int-tam-wei}
At the level of the effective-mass approximation, a hydrogenic donor impurity and the hydrogen 
atom are almost completely analogous quantum systems.\cite{koh}
Thus, it is in principle a simple matter to describe at that level the effect of the so-called 
extrinsic or Rashba spin-orbit interaction in the impurity states.
This contribution to the spin-orbit interaction in local external potentials in solids is, 
in this context, analogous to the spin-orbit coupling in the hydrogen atom.
However, care must be taken due to the necessary renormalization of the spin-orbit coupling 
constant and to possible modifications to the spin-orbit formula due to anisotropies of the 
crystal structure.
On the other hand, an additional contribution to the spin-orbit interaction, which is specific 
to the solid-state context must be taken into consideration.
This so-called intrinsic or Dresselhaus contribution\cite{dre} is present in the bulk of
the material and reappears in the envelope-function approximation theory of external, 
mesoscopic, potentials, like the one caused by a ionized donor.

In this article we calculate the energy levels of the $n=2$ shell of hydrogenic impurities
of semiconductors with wurtzite crystal structure in the presence of these two contributions 
to the spin-orbit interaction.
While, as mentioned above, the calculation of the extrinsic contribution is in a sense 
a straightforward application of the well-known formulas for the hydrogen atom, we need to
fill a theoretical gap, caused by the anisotropy of the wurtzite crystal structure.
To that effect, we obtain here an expression for the extrinsic spin-orbit interaction
in wurtzite semiconductors and the effective spin-orbit coupling constant for conduction
band electrons.
The calculation of the effects of the intrinsic spin-orbit interaction in this context
is entirely new and it is considered acting alone and in combination with the extrinsic
contribution.

The article is organized as follows.
In Section II we introduce the system and study the effect of the intrinsic spin-orbit interaction.
In Section III we derive the effective Hamiltonian of the extrinsic spin-orbit interaction using 
the Foldy-Wouthuysen transformation adapted to this context 
and in Section IV we calculate the energy corrections that it produces.
In Section V we study the combined effect of both spin-orbit interactions, and in Section VI we
provide the concluding remarks.


\section{Intrinsic spin-orbit interaction in the hydrogenic impurity}
\label{sec:efa-impurity}

We consider an electron bound to a hydrogenic donor impurity in a bulk semiconductor with
wurtzite crystal structure.
Working at the level of the envelope-function approximation (EFA), both the intrinsic 
and the extrinsic spin-orbit couplings appear in the Hamiltonian:
\begin{equation} 
  H = H_{\text{0}} + H_{\text{int}} + H_{\text{ext}},
\label{eq:Htot}
\end{equation}
where
\begin{equation} 
  H_0 = \frac{p^2}{2 m^*} + V(\mathbf{r}).
\label{eq:Hzero} 
\end{equation}
Here $V(\mathbf{r}) = - e^2/\epsilon r$ is the effective Coulomb potential of the electron
bound to the ionized donor impurity.
We assume the effective mass $m^*$ and the dielectric constant $\epsilon$ to be isotropic, 
thereby preserving the spherical symmetry of the hydrogenic Hamiltonian $H_0$.

The extrinsic spin-orbit coupling, $H_{\text{ext}}$, will be discussed below, and for the
moment we focus on the intrinsic contribution, $H_{\text{int}}$, which for semiconductors
with wurtzite crystal structure is given by \cite{voo-wil-car, fu-wu,wan-wu-tsa}
\begin{equation}
  H_{\text{int}} = \alpha \,( \sigma_x k_y  - \, \sigma_y k_x) 
                   + \gamma \,(b \,k_z^{2}-\,  k_{\parallel}^2)
                              ( \sigma_x k_y  - \, \sigma_y k_x),
\label{eq:int_completa}
\end{equation}
where $k_{\parallel}^2=k_x^2+k_y^2$ and $\alpha$, $b$, and $ \gamma$ are material-dependent 
parameters which are obtained experimentally or via ab-initio calculations.
$\boldsymbol{\hat{\sigma}}=(\sigma_x,\sigma_y,\sigma_z)$ are the Pauli matrices.
While $\alpha$ and $\gamma$ can vary considerably between different materials,
$b$ is roughly universal and close to 4 for all materials.
Note that $H_{\text{int}}$ has two parts, one of them linear and the other one cubic 
in the wavevector $k$.
The cubic-in-k term displays an anisotropy between the $z$-direction and the directions 
in the $xy$-plane.
This anisotropy and the presence of the linear term distinguish the intrinsic spin-orbit
Hamiltonian of wurtzite semiconductors from the Dresselhaus coupling of zincblende 
semiconductors.\cite{dre}

The hydrogenic Hamiltonian $H_0$ has the renormalized eigenvalues $E_n^*=-E_R^*/n^2$,
where $E_R^*=m^{\ast} e^4/2\epsilon^2\hbar^2$ is the effective Rydberg energy.
The aim of this study is to obtain the corrections to the $E_2$ energy level due to
the intrinsic and extrinsic spin-orbit Hamiltonians.
We will work at the level of first-order perturbation theory, which is adequate due
to the smallness of the spin-orbit couplings compared to the separation of the 
bare $E_n^*$ levels.
In order to diagonalize the intrinsic Hamiltonian $H_{\text{int}}$ in the $E_2$ subspace
we use the basis of hydrogenic eigenstates of $\{L^2,L_z,S^2,S_z\}$, given by
\begin{eqnarray}
  &&  \psi_{200 \eta} = \left( \frac{1}{32\pi a^{*3}} \right)^{\frac{1}{2}}
                        \left(2 - \frac{r}{a^*} \right) 
                        e^{-r/2a^*} |\eta\rangle     \nonumber \\
  &&  \psi_{211 \eta} = \left( \frac{1}{64\pi a^{*3}} \right)^{\frac{1}{2}} 
                        \frac{(x + i y)}{a^*} \,
                        e^{-r/2a^*} |\eta\rangle  \nonumber \\
  &&  \psi_{210 \eta} = \left( \frac{1}{32\pi a^{*3}} \right)^{\frac{1}{2}} \frac{z}{a^*} \,
                        e^{-r/2a^*} |\eta\rangle        \nonumber \\
  &&  \psi_{21-1 \eta} = -\left( \frac{1}{64\pi a^{*3}} \right)^{\frac{1}{2}} 
                         \frac{(x - i y)}{a^*} \,
                         e^{-r/2a^*} |\eta\rangle ,
\end{eqnarray}
where $\eta= \{\uparrow, \downarrow\}$.
The matrix elements of the linear-in-$k$ terms of $H_{\text{int}}$ in this basis are zero;
only the cubic-in-$k$ terms contribute.
Ordering the basis states as: 
$|200\uparrow\rangle$, $|200\downarrow\rangle$, 
$|211\uparrow\rangle$, $|211\downarrow\rangle$, $|210\uparrow\rangle$, 
$|210\downarrow\rangle$, $|21,-1\uparrow\rangle$, and $|21,-1\downarrow\rangle$, 
the matrix of $H_{\text{int}}$ in the $n=2$ subspace is
\begin{equation}
  \bar{\bar H}_{\text{int}}   \,=  \left(\begin{matrix}  
                                            
                         0   &   0   &   0   &   A   &   0   &   0   &    0   &   0 \\
                       
                         0   &   0   &   0   &   0   &   0   &   0   &    B   &   0  \\
                       
                         0   &   0   &   0   &   0   &   0   &   0   &    0    &   0  \\
                        
                         A   &   0   &   0   &   0   &   0   &   0   &    0    &   0\\
                         
                         0   &   0   &   0   &   0   &   0   &   0   &     0   &   0\\
                        
                         0   &   0   &   0   &   0   &   0   &   0   &     0   &   0  \\
                        
                         0   &   B   &   0   &   0   &   0   &   0   &     0   &   0  \\
                        
                         0   &   0   &   0   &   0   &   0   &   0   &     0   &   0  \\
                                                                       \end{matrix}\right) ,
\end{equation}
where
\begin{eqnarray}
  && A  = \frac{\gamma}{32\sqrt2 \, a^{*3}} \left(\frac{14}{15} b + \frac{133}{60} \right) \nonumber \\
  && B = \frac{\gamma}{32\sqrt2 \, a^{*3}} \left(\frac{62}{15} b + \frac{433}{60} \right) .
  \label{eq:AB}
\end{eqnarray} 
The secular equation for $\bar{\bar H}_{\text{int}}$, 
$\text{det} \left(\bar{\bar H}_{\text{int}} - \varepsilon \bar{\bar I} \right) = 0 $, 
yields
\begin{equation}
   \varepsilon^{4} - \varepsilon^{2}(A^2 + B^2) + (A\,\, B)^2 = 0 .
   \label{eq:intAB}
\end{equation}
The eigenvalues are then $\varepsilon_{1,2} = \pm A $ and $\varepsilon_{3,4} = \pm B $.
The other four eigenvalues are degenerate and equal to zero.
In Table I we present the non-zero energy corrections for the materials GaN, ZnO, InN and AlN, 
along with their $\gamma$ and $b$ parameters.
In the last two columns we present the energy splittings $2A$ and $2B$ as percentages
of the unperturbed energy $E_2^*$.

%
%
%
%
%
\begin{table}[ht]
\label{tab:int}
\begin{center}
\begin{tabular} {| l | c | c | c | c | c| c | c |}
\hline
      &  $\gamma $   &  $ b $ &  $ \varepsilon_{1,2} $  & $ \varepsilon_{3,4} $  & $ E_2^{*} $ & $2A/E_2^{\ast} $ & $2B/E_2^{\ast} $ \\
      &[$\text{meV} \text{\AA}^3$] &   &    [ $\mu$eV ]      &      [ $\mu$eV ]       &   [meV]     &     $ [\%] $     &     $ [\%] $   \\
   \hline 
    GaN  &    400    &  3.954   &   $\pm$ 13.24           &     $\pm$  52.79       &    11.97    &     0.22    &    0.88   \\
   \hline
    ZnO  &    320    &  3.855   &   $\pm$ 14.67           &     $\pm$  58.39       &    14.65    &     0.20    &    0.80   \\
   \hline
    InN  &    345    &  4.885   &   $\pm$ 14.71           &     $\pm$  59.49       &    16.20    &     0.18    &    0.73   \\
   \hline 
    AlN  &    6.45   &  3.767   &   $\pm$ 3.98            &     $\pm$  15.81       &    70.86    &     0.011   &    0.15  \\ 
    \hline
\end{tabular}
\end{center}
\caption{Intrinsic spin-orbit interaction corrections to the $E_2^*$ energy level of hydrogenic 
donor impurities in various semiconductors with wurtzite crystal structure.
The parameters $\gamma$ and $b$ are also indicated.
The effective masses and dielectric constants used to calculate $E_2^{*}$ are given in Table II.}
\end{table}

\section{Derivation of the Extrinsic spin-orbit interaction}
\label{cap:wurtzita-ext}

The spin-orbit Hamiltonian of an electron in vacuum in the presence of an electrostatic
potential $V_0(\mathbf{r})$ is given by
\begin{equation}
  H_{\text{so}} = \lambda \,\boldsymbol{\hat{\sigma}} \cdot 
                                         \mathbf{k}\times \nabla V_{0}(\mathbf{r}).
\end{equation}
When the electron is immersed in a semiconductor in the presence of a mesoscopic potential 
$V(\mathbf{r})$, the effective extrinsic spin-orbit Hamiltonian takes the form 
\begin{equation}
  H_{\text{ext}} = \lambda^{\ast} \, \boldsymbol{\hat{\sigma}} \cdot 
                             \mathbf{k} \times \boldsymbol{\nabla}V(\mathbf{r}),
\label{eq:HSOeff}
\end{equation}
where $\lambda^{\ast}$ is an effective coupling constant.
This expression is valid for semiconductors with zincblende crystal structure,
which presents a basic cubic symmetry.
The wurtzite crystal structure has less symmetry than the zincblende, due to the special role of
its c-axis.
This lack of isotropy is also present in the intrinsic spin-orbit coupling given above, 
Eq.\ \eqref{eq:int_completa}.
In what follows we shall derive an expression analogous to Eq.\ \eqref{eq:HSOeff} for semiconductors 
with wurtzite crystal structure.

We start with the $\mathbf{k} \cdot \mathbf{p}$ crystal Hamiltonian: 
\begin{equation}
\label{eq:kdotp}
  \mathcal{H} = \mathcal{H}_0 + \mathcal{H}_{\mathbf{k} \cdot \mathbf{p}} + \mathcal{H}_{\text{so}}
\end{equation}
where 
\begin{eqnarray}
  \mathcal{H}_0 &=& \frac{P^2}{2 m} + U(r) , \nonumber \\
  \mathcal{H}_{\mathbf{k} \cdot \mathbf{p}} &=& \frac{\hbar}{m}\mathbf{k}\cdot \mathbf{p} + 
      \frac{\hbar^{2} k^{2}}{2\,m} , \nonumber \\
  \mathcal{H}_{\text{so}} &=& \frac{\lambda}{\hbar} \, \boldsymbol{\hat{\sigma}} \cdot 
              \mathbf{p} \times \boldsymbol{\nabla}U ,
\end{eqnarray}
and $U$ is the periodic crystal potential.
We will write the matrix of $\mathcal{H}$ in the common basis of $H_0$ and $J_z$ given by:
\begin{eqnarray}
  &&  v_1 = \arrowvert i S \uparrow \rangle                                     \nonumber \\
  &&  v_2 = \arrowvert i S \downarrow \rangle                                   \nonumber \\
  &&  v_3 = -\frac{1}{\sqrt2} \left \arrowvert (X+iY) \uparrow \right\rangle    \nonumber \\
  &&  v_4 = -\frac{1}{\sqrt6} \Big[ \arrowvert (X+iY)
                     \downarrow\rangle -2\arrowvert Z \uparrow \rangle \Big]    \nonumber \\
  &&  v_5 = \frac{1}{\sqrt6} \Big[ \arrowvert (X-iY) \uparrow\rangle 
                    +2\arrowvert Z \downarrow \rangle \Big]                     \nonumber \\
  &&  v_6 = \frac{1}{\sqrt2}\arrowvert (X-iY) \downarrow\rangle                 \nonumber \\
  &&  v_7 = -\frac{1}{\sqrt3} \Big[ \arrowvert (X+iY) \downarrow\rangle
                    +\arrowvert Z \uparrow \rangle \Big]                        \nonumber \\
  &&  v_8 = -\frac{1}{\sqrt3}\Big[ \arrowvert (X-iY) \uparrow\rangle 
                    -\arrowvert Z \downarrow \rangle \Big] . 
\end{eqnarray}
Here $\arrowvert S \eta \rangle$ are conduction-band $s$-states, with energy $E_c$, and 
$\arrowvert X \eta \rangle$,  $\arrowvert Y \eta \rangle$, and $\arrowvert Z \eta \rangle$ 
are valence-band $p$-type states, with energy $E_v$.
The energy gap is given by $E_g = E_c-E_v$.
We calculate the matrix elements 
$\mathcal{H}_{ij} = \langle v_i\,\arrowvert \mathcal{H} \arrowvert v_j\,\rangle$,
where $ \{i,j = 1, \ldots, 8 \} $, and obtain

\begin{widetext}
\begin{equation}
  \mathcal{H} = \left(\begin{matrix}  
     E_c   &    0   &  \frac{-1}{\sqrt2}P_2k_+  & \sqrt{\frac{2}{3}} P_1k_z   &   \frac{1}{\sqrt6}P_2k_-   &   0   &  \frac{-1}{\sqrt3}P_1k_z   &   \frac{-1}{\sqrt3}P_2k_-  \\
                       
     0     &   E_c   &   0   &   \frac{-1}{\sqrt6}P_2k_+   &   \sqrt{\frac{2}{3}} P_1k_z   &   \frac{1}{\sqrt2}P_2k_-   &   \frac{-1}{\sqrt3}P_2k_+   &   \frac{1}{\sqrt3}P_1k_z  \\
                       
     \frac{-1}{\sqrt2}P_2k_-   &   0   &   E_v    &   0   &   0   &   0   &   0   &   0  \\
                        
     \sqrt{\frac{2}{3}} P_1k_z   & \frac{-1}{\sqrt6}P_2k_-  &   0   &   E_v   &   0   &   0   &  0  & 0  \\
                        
     \frac{1}{\sqrt6}P_2k_+   &   \sqrt{\frac{2}{3}} P_1k_z   &   0   &   0  &   E_v   &   0   & 0  \\
                   
     0   &   \frac{1}{\sqrt2}P_2k_+   &   0   &   0   &   0   &   E_v   &   0   &   0  \\
                        
     \frac{-1}{\sqrt3}P_1k_z    &    \frac{-1}{\sqrt3}P_2k_-   &   0   &   0  & 0  & 0 & E_v- \Delta_0    &   0  \\
                        
     \frac{-1}{\sqrt3}P_2k_+   &   \frac{1}{\sqrt3}P_1k_z   &   0   &   0   &   0   &   0   &   0   &   E_v -\Delta_0  \\
          \end{matrix}\right)
\end{equation}
\end{widetext}
where $k_{\pm} = k_x \pm ik_y$ and 
$\Delta_0 =  \frac{\hbar}{4 m^2 c^4} \langle X | \frac{\partial U}{\partial x} P_y - 
\frac{\partial U}{\partial y} P_x | Y \rangle$ is the spin-orbit splitting of the valence bands.
We have defined the constants $P_1$ and $P_2$ coming from the matrix elements:
\begin{equation}
 \frac{\hbar}{m}\langle -iS\,\downarrow\arrowvert\mathbf{k} \cdot \mathbf{p} 
   \arrowvert Z\,\downarrow\rangle\,=  \,-i \frac{\hbar}{m} k_z 
   \langle S\,\arrowvert p_z \arrowvert Z\rangle \equiv k_z P_1 , \nonumber
\end{equation}
\begin{equation}
 \frac{\hbar}{m}\langle -iS\,\downarrow\arrowvert\mathbf{k} \cdot \mathbf{p} 
    \arrowvert X\,\downarrow\rangle\,=  \, -i \frac{\hbar}{m} k_x 
    \langle S\,\arrowvert p_x \arrowvert X\rangle \equiv k_x P_2 .
\end{equation}

We now introduce the impurity potential $V(\mathbf{r})$, which varies slowly in the length 
scale of the lattice constant.
Its matrix elements in the basis \{$v_i$\} are essentially diagonal thanks to the orthogonality 
of the basis set and its slow variation in atomic scale.
In short, we are applying here the envelope function approximation. 
The matrix of $\mathcal{H}+V$ can be expressed in a compact form using the matrices $\mathbf{T}$ 
familiar from group theory:\cite{win}
\begin{equation}
 T_x  = \frac{1}{3\sqrt{2}}  
 \left(\begin{matrix} 
    -\sqrt{3} & 0 & 1 & 0\\
    0 & -1 & 0 & \sqrt{3}  
 \end{matrix}\right),
\end{equation}
\begin{equation}
 T_y  = \frac{-i}{3\sqrt{2}}  
 \left(\begin{matrix} 
    \sqrt{3} & 0 & 1 & 0\\
    0 & 1 & 0 & \sqrt{3}  
 \end{matrix}\right),
\end{equation}
\begin{equation}
 T_z  = \frac{\sqrt{2}}{3}  
 \left(\begin{matrix} 
    0 & 1 & 0 & 0 \\
    0 & 0 & 1 & 0  
 \end{matrix}\right).
\end{equation}
Using these matrices, the Hamiltonian matrix becomes:
\begin{equation}
 \left(\begin{matrix} 
   (E_c+V) \mathbb{I}_{2\times 2} & \sqrt3 P_1\mathbf{T} \cdot \mathbf{k}_{\alpha} & 
   \frac{-1}{\sqrt3}P_1 \boldsymbol{\hat{\sigma}} \cdot \mathbf{k_{\alpha}}      \\
                       
   \sqrt3 P_2\mathbf{T}^{\dagger} \cdot \mathbf{k}_{\alpha} & (E_v +V) \mathbb{I}_{4\times 4} & 0   \\
                       
   \frac{-1}{\sqrt3}P_1 \boldsymbol{\hat{\sigma}} \cdot \mathbf{k_{\alpha}} & 0 & 
   (E_v -\Delta_0+V) \mathbb{I}_{2\times 2} 
 \end{matrix}\right),
\end{equation}
where $\mathbf{k}_{\alpha}=(\alpha k_x, \alpha k_y, k_z)$ and $\alpha = P_2/P_1$.

Following the application of the Foldy-Wouthuysen transformation described by Winkler for 
zincblende semiconductors,\cite{win} we obtain an effective equation, restricted to the
conduction band, for the electronic states in the donor impurity
\begin{widetext}
\begin{equation}
  \left[ \mathbf{T} \cdot \mathbf{k}_{\alpha} \frac{3P_1^{2}}{E-V+E_g}\mathbf{T}^{\dagger} 
  \cdot \mathbf{k}_{\alpha} + \boldsymbol{\hat{\sigma}} \cdot \mathbf{k_{\alpha}}
  \frac{P^{2}_1}{3(E-V+E_g)}\boldsymbol{\hat{\sigma}} \cdot \mathbf{k_{\alpha}} \right]\psi_c=(E-V)\psi_c .
\end{equation}
\end{widetext}
Using the relation 
$ (\boldsymbol{\hat{\sigma}} \cdot \mathbf{A}) 
(\boldsymbol{\hat{\sigma}} \cdot \mathbf{B}) = \mathbf{A}\cdot \mathbf{B} + i\boldsymbol{\hat{\sigma}}
\cdot \left(\mathbf{A} \times \mathbf{B} \right)$, 
we obtain two terms from the second term in the above equation, one of which corresponds to the effective
spin-orbit interaction in the conduction band:
\begin{equation}
   H_{\text{ext}} = \lambda^{*}_w \, 
   \boldsymbol{\hat{\sigma}} \cdot \left(\mathbf{k_{\alpha}} \times \boldsymbol{\nabla}_{\alpha} V \right).
\label{eq:Hext}
\end{equation}
where we defined $\boldsymbol{\nabla}_{\alpha} \equiv (\alpha \frac{\partial}{\partial x}, 
\alpha \frac{\partial}{\partial y}, \frac{\partial}{\partial z} )$.
We have identified the coupling constant for the extrinsic spin-orbit interaction in wurtzite semiconductors:
\begin{equation}
   \lambda^{*}_w = \frac{\epsilon P_{1}^2}{3}\left[\frac{2}{E^{2}_g}  - 
   \frac{1}{(E_g +\Delta_0)^{2}} \right],
\label{eq:lambda}
\end{equation}
analogous to the known coupling constant $\lambda^{*}$ in Eq.\ \eqref{eq:HSOeff} for zincblende materials.

Note the factor $\alpha \equiv P_2/P_1$ in Eq.\ \eqref{eq:Hext}, which reflects the anisotropy
of the wurtzite crystal structure.
The Coulomb potential of the hydrogenic impurity, $V(\mathbf{r})$, that appears in Eq.\ \eqref{eq:Hext},
was introduced in Eq.\ \eqref{eq:Hzero}.
Actually, the spherically symmetric form given after Eq.\ \eqref{eq:Hzero} is a simplified expression 
which does not include the effect of the anisotropic effective mass and dielectric constant of wurtzite
crystal structures.\cite{han}
As a first approximation, here we will work with this spherically symmetric Coulomb potential
and will also disregard the $\alpha$-dependence of $\boldsymbol{\nabla}_{\alpha}$ and $\mathbf{k_{\alpha}}$.
%
A complete treatment of the anisotropy effects would require considering the modified
eigenvalue problem of the anisotropic hydrogenic impurity, and then the effect of the
factor $\alpha$ in the spin-orbit interaction.
We leave this refined treatment for future work.
In Table II we present the values of $\lambda^*_w$ for GaN, ZnO, InN, and AlN,
along with the material parameters needed to evaluate Eq.\ \eqref{eq:lambda}.

\begin{table}[ht]
\label{tab:lambda}
\begin{center}
\begin{tabular} {| l | c | c | c | c | c |}
\hline
      & $ m^*/m_0$      & $\epsilon$      &  $ E_g $  & $\Delta_0 $  &  $\lambda^{*}_w  $\\
      &                 &                 &   [eV]    &  [meV]       & [$10^{-2}$ \AA$^2$] \\
   \hline 
    GaN  & 0.32         &    9.53         &   3.51   &  72.9    &    5.95   \\
   \hline
    ZnO  & 0.32         &    8.62         &   3.44   &   43     &    3.08   \\
   \hline
    InN  & 0.26         &    7.39         &   0.78   &   40     &    1.33   \\
   \hline 
    AlN  & 0.38         &    4.27         &   5.4    &  -58.5   &   -1.04   \\ 
    \hline
\end{tabular}
\end{center}
\caption{Coupling constant of the effective extrinsic spin-orbit interaction 
 and auxiliary material parameters\cite{han} for wurtzite semiconductors.}
\end{table}


\section{Extrinsic spin-orbit corrections to the $2p$ level of hydrogenic impurities}

Using Eqs.\ \eqref{eq:Hext} and \eqref{eq:lambda} we obtain the Hamiltonian of the extrinsic spin-orbit 
interaction due to the Coulomb potential of the hydrogenic the donor impurity:
\begin{equation}
  H_{\text{ext}} =  \frac{24a^{*3}}{r^3} \xi_{2p}^{\ast}  \, \mathbf{L} \cdot \mathbf{S} ,
\label{eq:ext_final}
\end{equation}
where
\begin{equation}
   \xi^{\ast}_{2p} \equiv \frac{e^2 \lambda^{\ast}_w}{24 \epsilon \hbar^2 a^{*3}} .
\end{equation}
As anticipated above, we simplified the Hamiltonian $H_{\text{ext}}$ by setting the 
ratio $\alpha=1$ in Eq.\ \eqref{eq:Hext}.
We thus revert to the standard spin-orbit coupling of the hydrogen atom, but take into account
the appropriate, renormalized, coupling constant $\lambda^*_w$.
The calculation of the first-order corrections to the $2p$ energy level of the impurity due
to $H_{\text{ext}}$ now follows the standard treatment of spin-orbit interaction in the
hydrogen atom.
The common eigenvalues of $H_{\text{ext}}$ and $J^2$ are given by:
\begin{equation}
  \varepsilon_1 = \frac{1}{2}\xi_{2p}^{\ast} \Big[ \frac{3}{4}-2-\frac{3}{4} \Big]\hbar^2 = 
        - \xi_{2p}^{\ast} \hbar^2
\end{equation}
for $j=1/2$, and
\begin{equation}
  \varepsilon_2 = \frac{1}{2} \xi_{2p}^{\ast}  \Big[ \frac{15}{4}-2-\frac{3}{4} \Big]\hbar^2 =  
        \frac{1}{2} \xi_{2p}^{\ast} \hbar^2
\end{equation}
for $j=3/2$. 
We thus obtain for the energy corrections:
\begin{eqnarray}
  &&  \varepsilon_1 = - \frac{e^2 \lambda^{\ast}}{12 \epsilon a^{\ast 3}} \equiv -2 \,\beta     \nonumber \\
  &&  \varepsilon_2 =  \frac{e^2 \lambda^{\ast}}{24 \epsilon a^{\ast 3}}  \equiv \beta . 
  \label{eq:bett}
\end{eqnarray}
The numerical values of $\varepsilon_1$ and $\varepsilon_2$ are shown in Table III, 
together with the parameters needed for their evaluation.
We also give the energy variation as a percentage of the unperturbed energy,
$(\varepsilon_2 - \varepsilon_1) / E_2^{\ast}$.
One can see that the splitting due to extrinsic spin-orbit interaction is four orders of magnitude
smaller than the energy of the original level.
This ratio is small but it is not negligible as it is, in fact, one order of magnitude larger than 
the one obtained for the hydrogen atom, which is equal to 0.00133 \%.
It should be emphasized that this comparison between the hydrogenic impurity and the hydrogen atom
was not obvious a priori, since the renormalization of the coupling constant $\lambda$ is very 
pronounced (6 orders of magnitude) and could have produced results radically different.

We remark that for aluminum nitride (AlN) the relation between $\varepsilon_1$ and $\varepsilon_2$ 
is inverted.  
This peculiarity originates in the particular characteristics of its electronic structure, 
which cause $\lambda^{\ast}$ to become negative. 

\begin{table}[ht]
\label{tab:energias_SO_ext}
\begin{center}
\begin{tabular} {| l | c | c | c | c | c| c |}
\hline
      & $ a^{\ast} $  & $ \lambda^{\ast}_w$ &  $ \text{Ry}^{\ast} $  & $ \varepsilon_1 $  & $ \varepsilon_2 $ & $ (\varepsilon_2 - \varepsilon_1)/E_2^{\ast} $   \\
      & [\AA]         &[$10^{-2}$ \AA$^2$]&    [meV]         & [$\mu$eV]    & [$\mu$eV]  &    $ [\%] $      \\
   \hline 
    GaN  &    15.8    &      5.95         &    11.97         &     -1.94         &  0.968      &    0.024    \\
   \hline
    ZnO  &    14.1    &      3.08         &    14.65         &     -1.56         &  0.779      &    0.016    \\
   \hline
    InN  &    15.2    &      1.33         &    16.20         &     -0.625        &  0.313      &    0.0058   \\
   \hline 
    AlN  &    5.9     &     -1.04         &    70.85         &      14.50        & -7.25       &   -0.031    \\ 
    \hline
\end{tabular}
\end{center}
\caption{\footnotesize{Extrinsic spin-orbit corrections to the $2p$ energy level of hydrogenic 
donor impurities for four important wurtzite semiconductores, along with relevant material 
parameters.}}
\end{table}


\section{Combined intrinsic and extrinsic spin-orbit interactions}

We now study the effects of the intrinsic and extrinsic spin-orbit interactions combined on 
the $E_2^*$ energy level of hydrogenic impurities in wurtzite semiconductors.
Thus, we now consider the complete Hamiltonian, Eq.\ \eqref{eq:Htot}, with $H_{\text{int}}$
and $H_{\text{ext}}$ given in Eqs.\ \eqref{eq:int_completa} and Eq.\ \eqref{eq:ext_final},
respectively.
We will perform again a first-order perturbative treatment, now considering the full spin-orbit
Hamiltonian $H_{\text{int}}+H_{\text{ext}}$ as the perturbation.
To that end, we will express the spin-orbit coupling in the so-called uncoupled basis states 
of the $E_2^*$ shell, that is,  $\left\lbrace|l, s; m_l,m_s\rangle\right\rbrace$, used in 
Section \ref{sec:efa-impurity} to treat the intrinsic-alone case.
The matrix of the combined spin-orbit Hamiltonian in this basis is:
\begin{widetext}
\begin{equation}
  H_{\text{ext}} + H_{\text{int}} = \left(\begin{matrix}  
     -\varepsilon &     0    &       0      &     A         &      0         &       0      &      0         &      0   \\
                       
     0        & -\varepsilon &       0      &     0         &      0         &       0      &      B         &      0   \\
                        
     0        &    0     & \beta-\varepsilon&     0         &      0         &       0      &      0         &      0   \\
                        
     A        &    0     &       0      & -\beta-\varepsilon & \sqrt2\,\beta &       0      &       0        &      0   \\
                         
     0        &    0     &       0      &   \sqrt2\,\beta&  -\varepsilon     &       0      &       0        &      0   \\
                   
     0        &    0     &       0      &     0          &      0        &    -\varepsilon   & \sqrt2\,\beta &      0   \\
                         
     0        &    B     &       0      &     0          &      0        &     \sqrt2\,\beta & -\beta-\varepsilon &     0  \\
                        
     0        &    0     &       0      &     0          &      0        &       0        &      0       & \beta-\varepsilon   \\
 \end{matrix}\right) ,
\end{equation}
\end{widetext}
where $\beta$ was defined in Eq.\ \eqref{eq:bett} and $A$ and $B$ have been defined in 
Eqs.\ \eqref{eq:AB}. 

\begin{figure}
  \centerline{\includegraphics[scale=0.6]{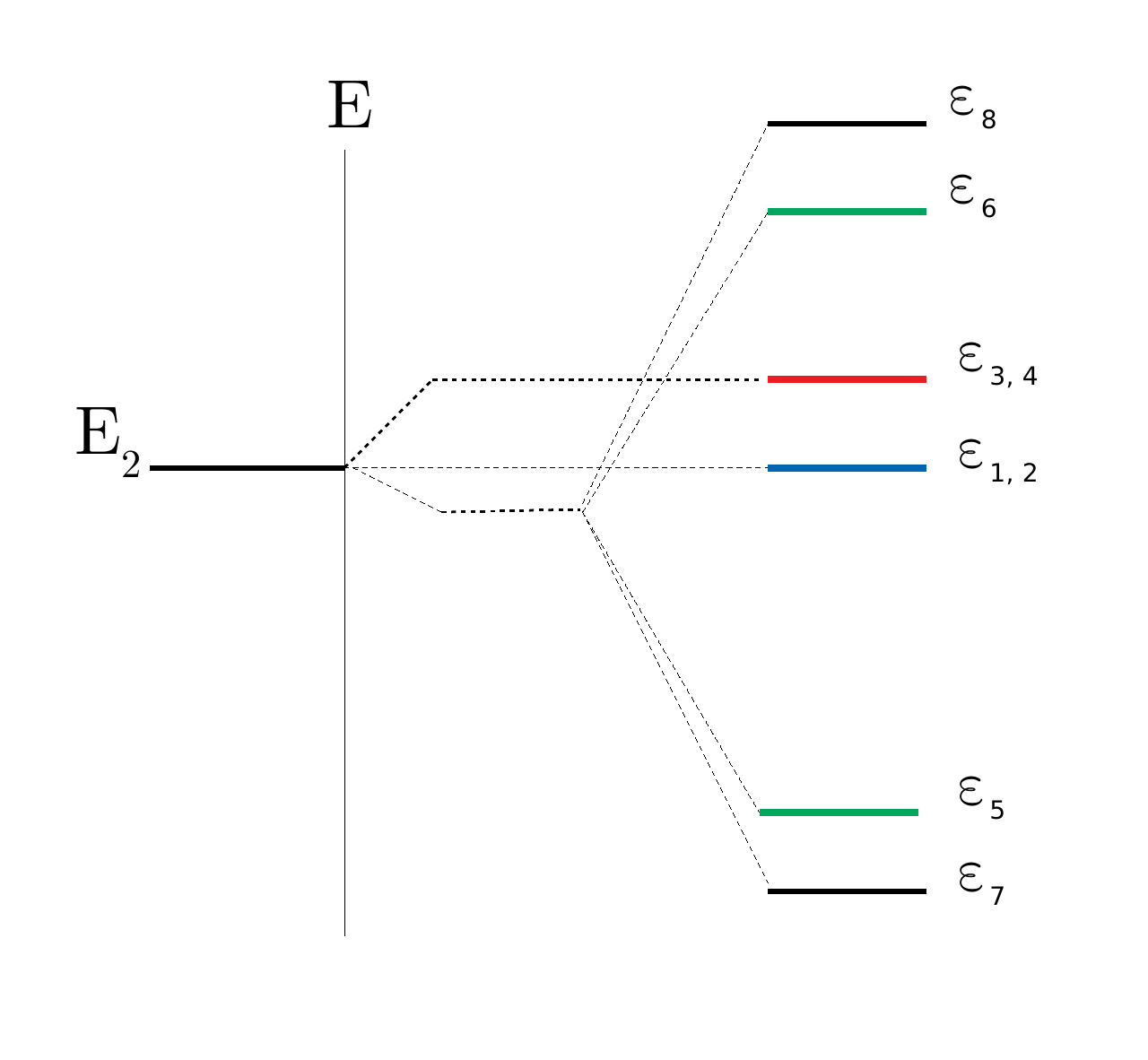}}
  \caption{\footnotesize{Schematic representation of the corrections to the energy level $E_2^*$ 
  of hydrogenic donor impurities to first order in the intrinsic and extrinsic spin-orbit 
  interactions combined.}
  \label{fig:combinado}}
\end{figure}

The characteristic polynomial that solves the eigenvalue problem is:
\begin{equation}
 \varepsilon^2 (\varepsilon-\beta)^2(\varepsilon\,\beta+\varepsilon^2-A^2-2\,\beta^2)
 (\varepsilon\,\beta+\varepsilon^2-B^2-2\,\beta^2)=0 ,
\end{equation}
and the corresponding eigenvalues are:
\begin{equation}
 \varepsilon^2 = 0  \,\,\,\,\,\,\,\,\,\,\,\, \Rightarrow  \,\,\,\,\,  \varepsilon_{1,2} = 0,  
 \label{eq:rot}
\end{equation}
\begin{equation}
 \,\,\,(\varepsilon-\beta)^2 = 0  \,\, \Rightarrow \,\,\,\,\,\,\, \varepsilon_{3,4} = \beta , 
 \label{eq:rot1}
\end{equation}
\begin{equation}
 \varepsilon^2+\varepsilon\beta-(A^2+2\beta^2) = 0  \,\,\,\,\,\,\,\,  \Rightarrow  \,\,\,\,\, \varepsilon_{5,6} = 
 -\frac{\beta}{2} \pm \sqrt{\frac{9}{4}\beta^2 + A^2},  
 \label{eq:rot2}
\end{equation}
\begin{equation}
 \varepsilon^2+\varepsilon\beta-(B^2+2\beta^2) = 0  \,\,\,\,\,\, \Rightarrow  \,\,\,\,\, \varepsilon_{7,8} = 
 -\frac{\beta}{2} \pm \sqrt{ \frac{9}{4} \beta^2 + B^2}.
  \label{eq:rot3}
\end{equation}
We remark that the energy corrections given in Eqs.\ \eqref{eq:rot}--\eqref{eq:rot3} contain the 
previous cases (intrinsic and extrinsic spin-orbit interactions acting alone) in the appropriate 
limits, and they are represented schematically in Fig. \ref{fig:combinado}.
This plot offers a qualitative view of the energy splittings and shows the greater breaking of
degeneracy caused by the combined action of both spin-orbit couplings.
Finally, we calculate the eigenvalues $\varepsilon_i$ ($i=3, \ldots ,8$) using the values 
of $A$, $B$ y $\beta$ corresponding to GaN, ZnO, InN y AlN; the results are given in Table IV.

\begin{table}[ht]
\label{tab:comb}
\begin{center}
\begin{tabular} {| l | c | c | c | c | c |}
\hline
      & $\varepsilon_{3,4}$ & $\varepsilon_5$ & $\varepsilon_6$  & $\varepsilon_7$ & $\varepsilon_8$ \\
      &     [$\mu$eV]    &      [$\mu$eV]     &    [$\mu$eV]    &   [$\mu$eV]      &    [$\mu$eV]    \\
   \hline 
    GaN &    0.23       &      -13.36        &    13.13         &    -54.71        &    54.48     \\
   \hline
    ZnO  &    4.90      &      -49.39        &     44.94        &    -60.78        &    56.33    \\
   \hline
    InN  &    0.76      &      -15.42        &    14.66         &     -61.31       &    60.63     \\
   \hline 
    AlN  &   -20.15     &      -20.44        &    40.58         &    -17.56        &   37.70      \\ 
    \hline
\end{tabular}

\caption{\footnotesize{Energy corrections due to the combined intrinsic and extrinsic spin-orbit
interactions to the energy level $E_2^*$ of hydrogenic donor impurities for four important  
binary semiconductors with wurtzite crystal structure.}}
\end{center}
\end{table}
%


\section{Conclusion}
\label{sec:conclusion}

We have studied theoretically the effects of the spin-orbit interaction on the
$E_2^*$ energy level of hydrogenic donor impurities embedded in semiconductors
with wurtzite crystal structure.
Both the intrinsic (Dresselhaus) and extrinsic (Rashba) spin-orbit interations
have been considered, first acting separately and then together.
The study was carried out at the level of first-order perturbation theory,
which turns out to be appropriate given the relative magnitude of the corrections
to the unperturbed energy spacings.
Furthermore, in order to evaluate the extrinsic spin-orbit interaction 
it was necessary to calculate the renormalized coupling constant $\lambda^*$ 
for wurtzite semiconductors from 8-band Kane theory.

We applied our calculations to four currently important semiconductors, i.e.\ GaN, ZnO 
InN, and AlN.
A general conclusion of these calculations is that both spin-orbit couplings produce
relative energy corrections that are bigger than the standard spin-orbit corrections
to the $E_2$ energy level of the hydrogen atom.
While for  GaN, ZnO, InN we conclude that the intrinsic spin-orbit interaction
produces larger energy corrections than the extrinsic one, that is not the case for AlN,
where both interactions have comparable effects.
Another anomaly shown by AlN is the fact that its effective coupling constant 
$\lambda^*$ is negative.
This causes the eigenvalues $\varepsilon_{3,4}$, which are positive for GaN, ZnO 
and InN, to become negative for AlN.
These anomalies of AlN are due to the specific features of its electronic structure
which determine the relevant parameters $\Delta_0$, $\gamma_w$ and $a^{\ast}$.
Finally, we have found that the combined action of both types of spin-orbit coupling 
leads to an almost complete breaking of the degeneracy of the unperturbed energy level.

\bigskip

 
\acknowledgments
We thank Rodolfo Jalabert and Dietmar Weinmann for discussions that motivated this work.
We gratefully acknowledge financial support from Projects UBACYT and CONICET PIP.


\end{document}